\documentclass[
 reprint,
superscriptaddress,
 amsmath,amssymb,
 aps,
superscriptaddress,
 amsmath,amssymb,
 aps,longbibliography
]{revtex4-1}

\usepackage{graphicx}
\usepackage{dcolumn}
\usepackage{bm}
\usepackage{amsmath,amssymb,amstext,amsthm}
\usepackage{braket}
\usepackage{bbold}
\usepackage{bbm}
\usepackage{xcolor}
\usepackage{ gensymb }
\usepackage{soul}
\usepackage[utf8]{inputenc}
\usepackage{booktabs}
\usepackage{comment}
\setstcolor{red}

\begin{document}

\preprint{APS/123-QED}
\setlength{\abovedisplayskip}{1pt}
\title{A generalized K-space coherent averaging method for engineering lattices of spin-orbit beams}

\author{Pinki Chahal}
\email{pinki@buffalo.edu}
\affiliation{Department of Physics, University at Buffalo, State University of New York, Buffalo, New York 14260, USA}

\author{Naume Shentevski}

\affiliation{Department of Physics, University at Buffalo, State University of New York, Buffalo, New York 14260, USA}

\author{Priyanka Vadnere}

\affiliation{Department of Physics, University at Buffalo, State University of New York, Buffalo, New York 14260, USA}

\author{David G. Cory}
\affiliation{Institute for Quantum Computing, University of Waterloo,  Waterloo, ON, Canada, N2L3G1}
\affiliation{Department of Chemistry, University of Waterloo, Waterloo, ON, Canada, N2L3G1}

\author{Owen Lailey} 
\affiliation{Institute for Quantum Computing, University of Waterloo,  Waterloo, ON, Canada, N2L3G1}
\affiliation{Department of Physics and Astronomy, University of Waterloo, Waterloo, ON, Canada, N2L3G1}

\author{Dmitry A. Pushin}

\affiliation{Institute for Quantum Computing, University of Waterloo,  Waterloo, ON, Canada, N2L3G1}
\affiliation{Department of Physics and Astronomy, University of Waterloo, Waterloo, ON, Canada, N2L3G1}

\author{Dusan Sarenac}
\email{dusansar@buffalo.edu}

\affiliation{Department of Physics, University at Buffalo, State University of New York, Buffalo, New York 14260, USA}

\date{\today}

\begin{abstract}

 Spin-orbit beams, in which the orbital angular momentum degree of freedom is coupled to a two-level system such as polarization of light or spin in electrons and neutrons, have gained significant interest for their unique propagation properties and potential applications in imaging, material characterization, optical trapping, and quantum information processing. In this work we introduce a method for generating and engineering two-dimensional lattices of such spin-orbit beams based on coherent averaging in k-space. By programming the angle, amplitude, and polarization of a set of input beams we obtain precise control over lattice geometry and period, as well as the orbital and radial degrees of freedom inside each unit cell. We explore both electromagnetic and matter wave implementations, and we experimentally demonstrate the generation and characterization of a micron-scale optical hexagonal lattice with well defined orbital and radial numbers in each unit cell. The described methods provide a robust and general method of generating and controlling structured waves such as optical skyrmions and matter wave implementations of orbit and spin-orbit beams.

\end{abstract}

\pacs{Valid PACS appear here}


\maketitle

\section{Introduction:}

Since the early 1990s, studies on orbital angular momentum (OAM) have advanced significantly and are now included in a broader framework of structured waves ~\cite{barnett2017optical, rubinsztein2016roadmap, bliokh2023roadmap, shen2019optical}. OAM has been demonstrated in beams of light ~\cite{allen1992orbital}, electrons~\cite{uchida2010generation, bliokh2007semiclassical}, and neutrons ~\cite{sarenac2022experimental, sarenac2024small}. It is also possible to create structured waves called spin-orbit beams (or vector-vortex beams) that exhibit coupling between polarization and OAM ~\cite{maurer2007tailoring,marrucci2006optical} in the case of photons, while for electrons and neutrons, the correlations are between spin and OAM~\cite{karimi2012spin, nsofini2016spin,sarenac2018methods}. 
Complex two- and three-dimensional (2D and 3D) lattices of spin–orbit states have attracted considerable interest due to their potential applications ranging from microscopy and micromanipulation in biology~\cite{liu2022super,curtis2002dynamic} to materials processing~\cite{gao2018quasi,baltrukonis2021high} and advanced optical functionalities such as trapping, sorting, and beam structuring~\cite{yang2021optical,guo2010optical,wei2009generation}.

Various experimental techniques have been introduced to generate such lattices~\cite{tsesses2018optical, du2019deep,gutierrez2021optical, sarenac2018generation, ruchi2017generation,cameron2021remote}. A common method involves using a spatial light modulator (SLM), which can manipulate various properties of light, such as amplitude, phase, and polarization ~\cite{curtis2002dynamic}. However, generating similar lattices in matter waves (electrons and neutrons) requires different approaches ~\cite{sarenac2019generation,geerits2023phase,verbeeck2010production} due to the lack of analogous devices to the SLM. One optics technique that has the potential for general and wide implementation is multi-beam interference (MBI), also known as interferometric lithography ~\cite{burrow2011multi}. Traditionally, MBI techniques have primarily enabled control over lattice geometry, such as symmetry and periodicity~\cite{wu2006fabrication,stay2009conditions,zhang2023hexagonal, stay2008contrast,shoji2006multi,li2022generation,vyas2007vortex,cai2001formation}. However, there is a need for a general method that also allows robust and systematic control over unit‑cell parameters, including OAM number ($\ell$) and radial number ($n_r$).

In this work, we generalize the MBI technique by introducing a k-space coherent averaging approach to engineer lattices of spin-orbit beams of the desired topological properties, applicable to both electromagnetic and matter-wave systems. The lattices are generated by polarization encoding a set of input beams corresponding to the Fourier spectrum of the lattice geometry. Independent control is enabled over structural parameters, including period and lattice symmetry (e.g., square, hexagonal), as well as unit-cell parameters such as $\ell$ and $n_r$. We demonstrate the method both theoretically and experimentally by generating a micron-scale hexagonal lattices of optical spin-orbit beams, where the polarization structure at each site is engineered to exhibit desired OAM and radial characteristics.  Lastly, we examine a proposed design for matter-wave systems that is implementable with current experimental capabilities.   

\begin{table*}[t]
\centering
\renewcommand{\arraystretch}{1.3}
\begin{tabular}{c}
    \includegraphics[width=0.9\textwidth]{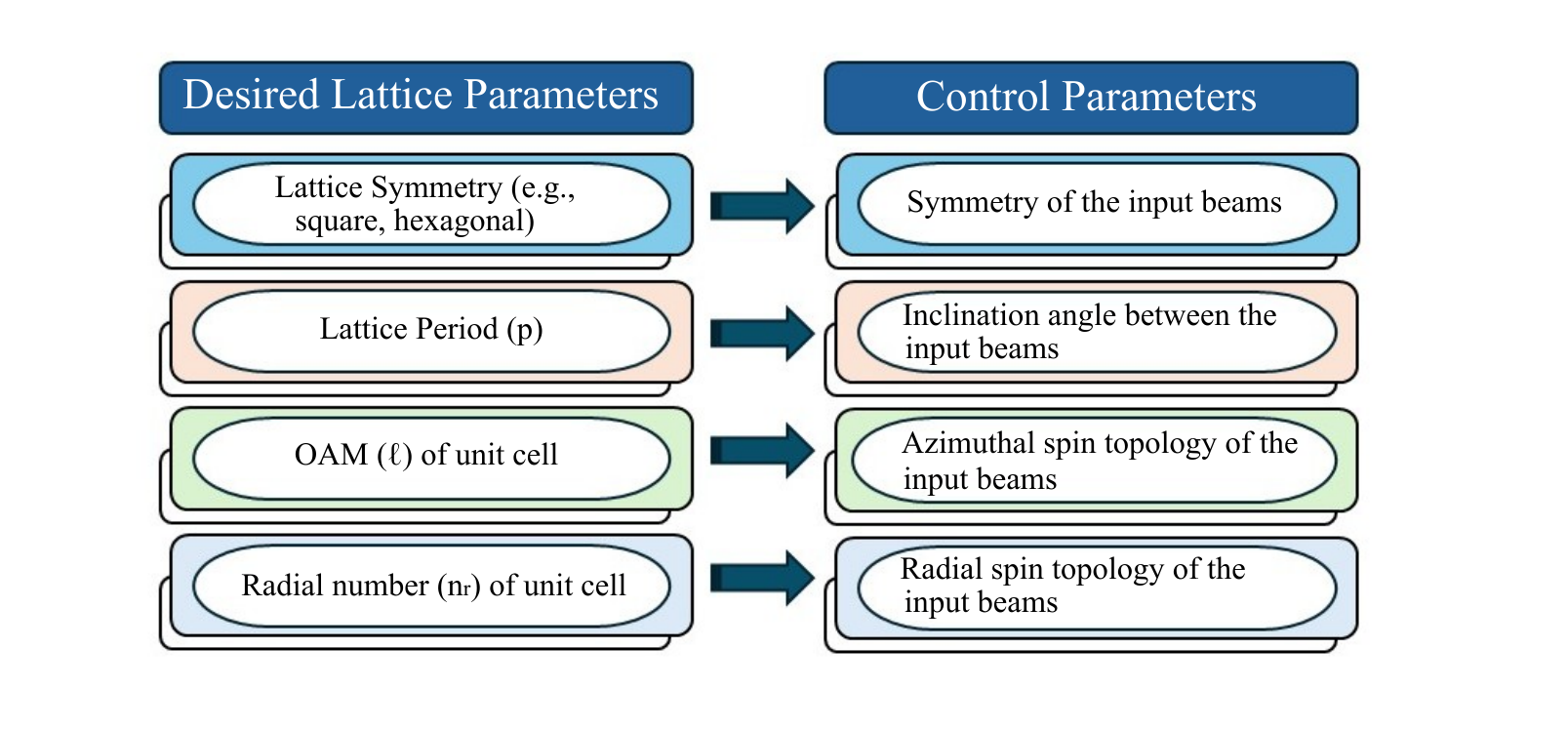} \\
\end{tabular}
\caption{Relation between the desired lattice parameters and their corresponding experimental control parameters. Spin–orbit features of the unit cell, such as $\ell$ and $n_r$, arise from the azimuthal and radial spin topology of the input beams, while the lattice symmetry and period are determined by the symmetry and inclination angle of the input beams.}
\label{tab:control_parameters}
\end{table*}

The technique described here pave the way for a novel imaging modality based on the analysis of spatial modes and beam topology. By analyzing changes in the polarization coupled OAM and radial number of specially prepared structured beams, and by tuning the lattice periodicity to match characteristic spatial structures in the sample, this approach may enable probes that are selectively sensitive to specific material properties. This would complement the impact that OAM has brought to imaging by providing enhanced phase contrast, higher sensitivity, access to super resolution ~\cite{shen2019optical,ritsch2017orbital,torner2005digital,zhang2016perfect,tan2010high} as well offering new scattering pathways and selection rules~\cite{schmiegelow2016transfer, afanasev2018experimental,sherwin2022scattering}.

\section{K-Space Coherent Averaging}
\noindent Table~\ref{tab:control_parameters} outlines the beam and unit-cell parameters accessible in the k-space coherent-averaging scheme, along with the experimental control parameters through which they can be independently controlled. The steps of the model can be summarized as follows: (1) determine the set of input beams with defined angular separation and amplitude, which sets the lattice symmetry; (2) assign each beam a polarization state, which sets the $\ell$ and $n_r$ in each unit cell; (3) The total wavefunction in k-space is obtained by the sum of these input beams:

\begin{figure*}
\centering\includegraphics[width=\linewidth]{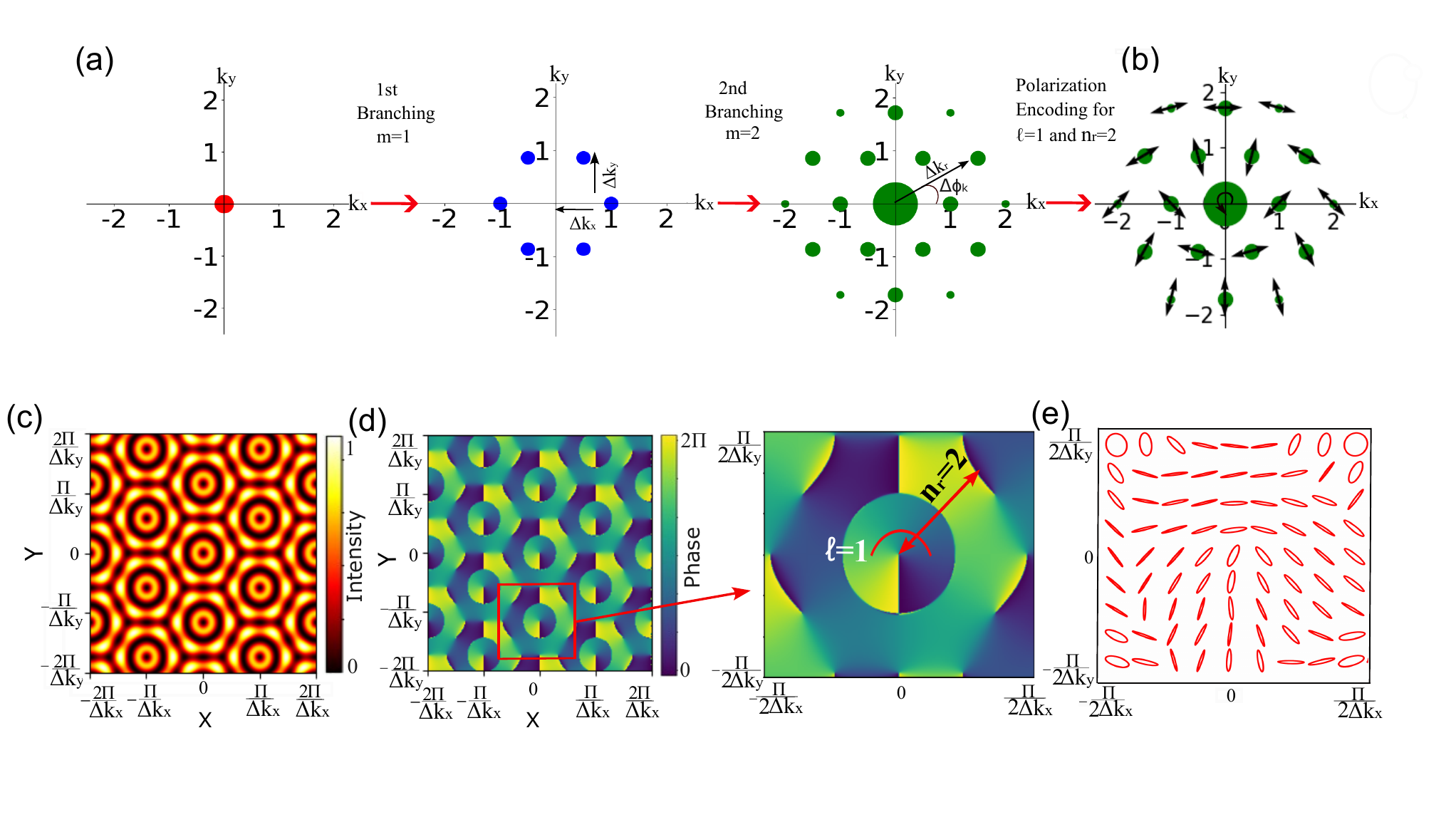}
\caption{Example of k-space coherent averaging to generate a hexagonal lattice with $\ell=1$ \& $n_r=2$. a) The first step of the lattice-generation method, where an initial on-axis input beam at (0,0) is transformed through successive applications of the branching operator \( \mathcal{B} \), producing six beams in the first iteration and 19 unique beams (out of 36 total) after the 2nd iteration of the \( \mathcal{B} \). Circle size indicates the relative amplitude. b) Polarization encoding of the 19 branched beams, where each beam is assigned a polarization direction corresponding to a spin–orbit state of $\ell=1$ and $n_r=2$. (c) Resulting real-space intensity profile, post-selected on left-circular polarization (LCP), obtained from the coherent superposition of the polarization-encoded beams in (b). (d) Corresponding phase distribution, and the inset shows a single lattice site from the phase map. The azimuthal phase winding of $2\pi$ illustrates $\ell=1$, while the presence of two radial phase reversals indicates the $n_r=2$. e) State of polarization distribution for a unit cell, showing how polarization varies spatially along the lattice site.}

\label{fig:fig1}
\end{figure*}

\begin{widetext} 
\begin{equation}
\Psi_{\text{out}}(k_x, k_y) = 
A(0, 0) \delta(k_x,k_y)\begin{bmatrix}
1 \\
0
\end{bmatrix}+\sum_{(\Delta k_x, \Delta k_y)} A(\Delta k_x, \Delta k_y) \delta(k_x-\Delta k_x,k_y-\Delta k_y) 
\begin{bmatrix}
1 \\
e^{i\ell \Delta \phi_k} \, \text{sgn}[\sin(\frac{n_r \pi}{m} \Delta k_r)]
\end{bmatrix}
\label{eq:psi_final}
\end{equation}
\end{widetext}

\noindent where for each input beam: the polarization state is represented by the 2 by 1 vector, $\ket{R} = \begin{pmatrix} 1 \\ 0 \end{pmatrix}$ and $\ket{L} = \begin{pmatrix} 0 \\ 1 \end{pmatrix}$ denote right- and left-circular polarization states respectively, $A(\Delta k_x, \Delta k_y)$ represents the amplitude, $\Delta k_x$ \& $\Delta k_y$ are the transverse wave vectors; they determine the angle of propagation between the beams thereby setting the lattice symmetry and period, $\Delta k_r$ \& $\Delta \phi_k$ are the corresponding radial and azimuthal wave vectors, and $m$ is the number of branching transformation iterations, described in the following section. The total wavefunction $\psi_{\text{out}}(x, y)$ in real space is obtained by the Fourier transform of $\Psi_{\text{out}}(k_x, k_y)$.

\subsection{Amplitude and Angular Separation of the Input Beams}
The coherent averaging process requires a set of input beams corresponding to the Fourier spectrum of the lattice geometry. The number of Fourier orders is determined by the target $\ell$ and $n_r$, while the corresponding amplitudes between the different orders must be appropriately defined. These amplitudes can be determined through a branching transformation method where the required set of beams is generated through the application of a branching operator \( \mathcal{B} \). Similar to a grating operator, \( \mathcal{B} \) acts on the wavefunction to coherently split it into multiple components directed along a set of displacement vectors, which define the angular separation and symmetry of the beams. Successive applications of \( \mathcal{B} \) yield the final set of beams. To illustrate this, consider the application of a branching operator onto some initial state $\Psi_1(k_{x1}, k_{y1})$, which yields $\Psi_2(k_{x2}, k_{y2})$:

\begin{equation}
\begin{split}
&\Psi_2(k_{x2}, k_{y2}) = \mathcal{B}\big[\Psi_1(k_{x1}, k_{y1})\big] \\
&= \sum_{(\Delta k_{x}, \Delta k_{y}) \in d} \frac{1}{s}\Psi(k_{x1}+\Delta k_x,k_{y1}+\Delta k_y)
\end{split}
\end{equation}

\noindent where $s$ is determined by the number of points in the lattice unit cell ($s=4~ (6)$ for square (hexagonal) lattice). The iterative application of \( \mathcal{B} \) leads to overlapping beams, which are combined through the addition of their amplitudes. The set \( d \) contains all relative displacement vectors that define the branching directions in the transverse plane for the desired symmetry. The overall choice of \( d \) defines the lattice geometry. See Fig.~\ref{fig:fig1}a for an example of the branching operation for a hexagonal symmetry. In the first iteration, the branching operator produces six distinct beams, each with a unique wave vector, and after the second iteration (m=2), 36 beams are generated, of which 19 are unique due to overlapping positions, where the size of each circle indicates the relative amplitude in the figure.

The resulting set of beams can be expressed as:

\begin{equation}
\Psi_{\text{out}}(k_x,k_y) = \mathcal{B}^m \left[ \Psi_{\text{1}}(k_{x1}, k_{y1}) \right],
\end{equation}

\noindent where the number of iterations \( m \) required to generate a lattice with a given \( \ell \) and \( n_r \) is given by the following relations. For a hexagonal lattice, the number of iterations satisfies $m\geq n_r$ \& $m \geq \ell/2.$ For a square lattice, the condition becomes $m\geq 2n_r$ \& $m\geq\ell$.

\begin{figure*}
\centering\includegraphics[width=\linewidth]{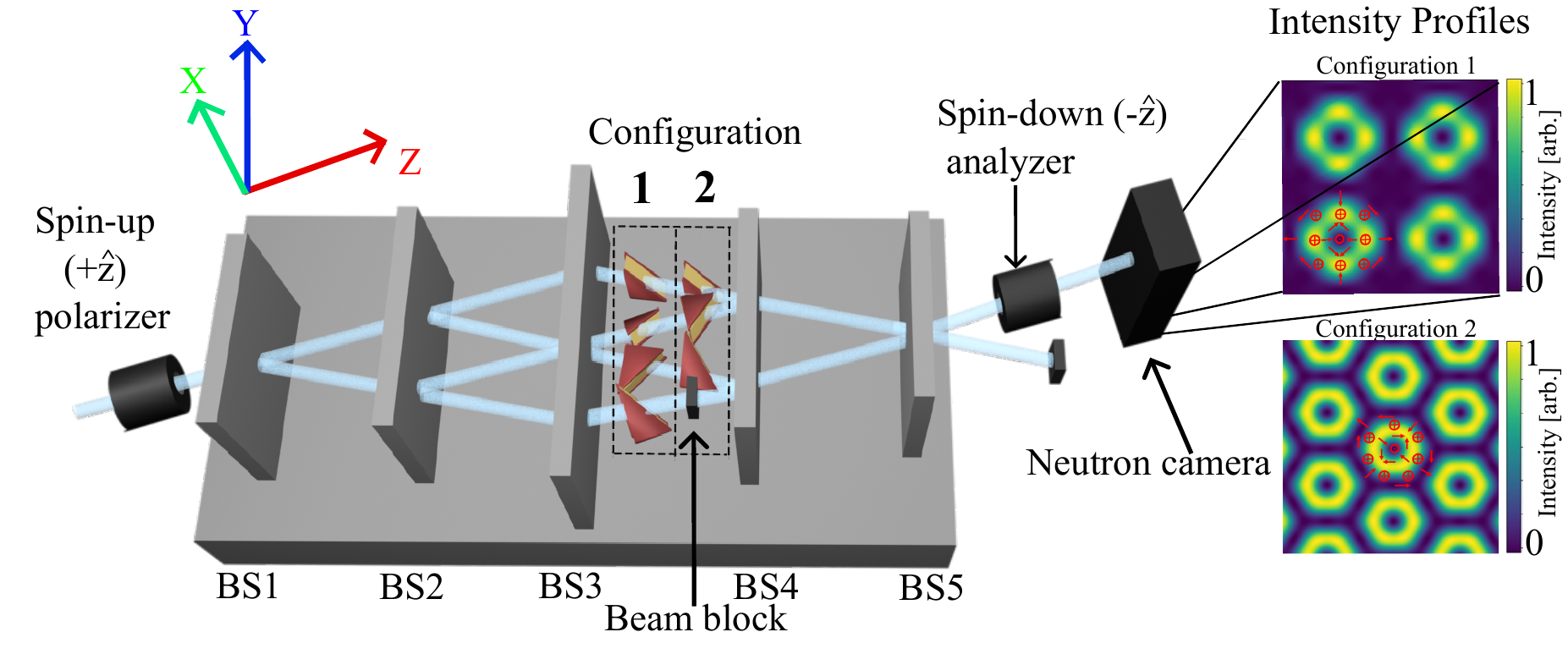}
\caption{ Schematic of the proposed setup for generating neutron spin–orbit lattices with orbital angular momentum $\ell=1$ and $n_r=1$ using a five-blade neutron interferometer (NI). An incident neutron beam is coherently split by the first beam splitter (BS1) into two paths, which are further divided by the second beam splitter (BS2) to form four distinct beams. Each of the four paths in Configuration 1 (for the square lattice) contains a magnetic prism oriented at $0^\circ$, $45^\circ$, $90^\circ$, and $135^\circ$, that imparts controlled angular deflection and spin rotation to encode the spin–orbit coupling. In Configuration 2, only three prisms are used, oriented at $0^\circ$, $60^\circ$ and $120^\circ$,  while the fourth path is blocked to produce the hexagonal lattice geometry. The beams are redirected and recombined by subsequent beam splitters (BS3–BS5), with the neutron's wave function forms a superposition of the four paths that constitutes a lattice of spin–orbit states. After the final recombination at BS5, the output beam is analyzed by a spin analyzer, and the resulting two-dimensional intensity distribution is recorded on a neutron detector.  
}

 \label{fig:fig2}
\end{figure*}

\subsection{Polarization Encoding}
The next step is to assign each beam a polarization state defined by the beam's location in k-space according to:

\begin{equation} 
\begin{bmatrix}
1 \\
e^{i\ell \Delta \phi_k} \operatorname{sgn} \left[ \sin\left( \frac{\pi n_r \Delta k_r}{m} \right) \right]
\end{bmatrix},
\end{equation}

\noindent where $\ell$ and $n_r$ are the desired OAM and radial numbers. Fig.~\ref{fig:fig1}b shows the polarization state assigned to each beam after the second iteration (m=2) of $\mathcal{B}$ for $\ell=1$ and $n_r=2$. Equation (6) assigns polarization as a function of the beam’s transverse momentum and is therefore applicable only for nonzero transverse wave vectors, $(k_x,k_y)\neq(0,0)$. Fig.~\ref{fig:fig1}c presents the polarization-dependent intensity profile of the desired lattice state obtained through the addition of the polarization-encoded beams shown in Fig.~\ref{fig:fig1}b, where the resulting lattice has a period inversely proportional to the angular separation of the input beams. The intensity profile shows two bright rings correspond to $n_r=2$. The phase map (Fig.~\ref{fig:fig1}d) demonstrates an OAM($\ell=1$) through a single $2\pi$ azimuthal phase winding. Fig.~\ref{fig:fig1}e illustrates the polarization state distribution of a unit cell, plotted using the Stokes parameters.

 \noindent

\section{Matter wave implementation using neutrons}

To apply our spin-orbit lattice generation technique to matter waves, we propose a practical approach for neutron beams. In particular, we outline a method for generating two-dimensional lattices of spin-orbit beams using a five-blade neutron interferometer (NI) \cite{nsofini2017noise} and neutron polarization optics elements, as shown in Fig.~\ref{fig:fig2}.

An incident monochromatic neutron beam, spin polarized along the propagation direction (z-axis), is directed into a multi-blade NI, where it is coherently split into four distinct paths through a sequence of beam splitters (BS1–BS2). The first beam splitter (BS1) divides the incoming wavefront into two coherent subbeams, which are further split by the second beam splitter (BS2), forming a four-path interferometric geometry. Each interferometric path may contain a magnetic prism to create the required k-space geometry of the lattice. Magnetic prisms are used, as they enable coupling between spin and transverse momentum, allowing each prism to split a neutron beam into two components corresponding to opposite spin projections and transverse momentum shifts. The operator associated with a magnetic prism that produces a magnetic field oriented along the unit vector~$\Vec{n}=(\sin{\alpha},\cos{\alpha},0)$ and whose incline face is oriented along the unit vector $\Vec{n}~'=(-\sin{\alpha},\cos{\alpha},0)$ perpendicular to $\vec{n}$, is given by:

\begin{equation}
    \hat{U}_{n} =
    \exp\Big[- i\frac{\pi}{a} \, 
    ({\vec{n}~' \cdot {\vec{r}}}) \, 
    (\hat{\vec{\sigma}} \cdot {\vec{n}}) 
    \Big]
    \label{eq:prism_operator}  
\end{equation}
where 
\begin{equation}
    a = \frac{2\pi h}{\gamma\left|B\right| m_n \lambda\tan(\beta)}
\end{equation}
 and $h$ is Planck's constant, $\gamma$ is the neutron gyromagnetic ratio, $B$ is the magnetic field strength, $m_n$ is the neutron mass, $\lambda$ is the neutron wavelength, $\beta$ is the incline angle of the prism and $\vec{r} = (x,y,z)$ is the position vector. 
 The incident neutron wavefunction, expressed in the transverse spin basis defined by the magnetic field vector direction $\vec{n}$, can be approximately written as: 
 
 \begin{equation}
    \psi_{in}\approx e^{i k_{0}z}\ket{\uparrow}_{z} = e^{ik_{z}z}\left(\frac{1}{\sqrt{2}}\ket{\uparrow}_{n}+\frac{1}{\sqrt{2}} e^{i\alpha}\ket{\downarrow}_{n}\right)
   \label{eq:spinor_n} 
\end{equation}

 \noindent where $k_0$ is the incoming neutron wavenumber, $\ket{\uparrow}_{n}$ and $\ket{\downarrow}_{n}$ are the eigenstates of $\hat{\vec{\sigma}} \cdot {\vec{n}}$.  The state resulting from the action of \eqref{eq:prism_operator} on \eqref{eq:spinor_n} is given by:
 
\begin{equation}
    \psi_{out} = e^{ik_{z}z}\left(e^{-i\vec{k}_\perp\cdot \vec{r}}\ket{\uparrow}_{n}+e^{i\alpha}e^{i\vec{k}_\perp\cdot\vec{r}}\ket{\downarrow}_{n}\right)
\label{eq:out_state}
\end{equation}

\noindent where $\vec{k_z}$ points along $\vec{z}$,   $\vec{k_\perp}$ points along $\vec{n'}$, and $|\vec{k}_\perp|=\frac{\pi}{a}$. Equation \eqref{eq:out_state} represents a superposition of two plane waves that are spin polarized along $\ket{\uparrow}_n$ and $\ket{\downarrow}_{n}$ respectively and propagating in slightly different directions. The angular separation between the two propagation directions is approximately:

\begin{equation}
    \theta\approx\frac{2k_\perp}{k_z}=\frac{\lambda^{2}\,m_n\gamma\left|B\right|}{2\pi h}\tan{(\beta)}
    \label{eq:splitting angle}
\end{equation}

\noindent To generate a square lattice, magnetic prisms are placed in all four interferometric paths (Configuration 1), with magnetic fields oriented at $0^\circ$, $45^\circ$, $90^\circ$, and $135^\circ$. The prisms at $0^\circ$ and $90^\circ$ produce a splitting angle $\theta$ given by equation $\ref{eq:splitting angle}$, while the prisms at $45^\circ$ and $135^\circ$ produce $\theta/\sqrt{2}$. The splitting angle can be tuned  by adjusting either the strength of the magnetic field $B$ or the prism incline angle $\beta$. The four prisms give rise to eight beam components whose transverse momentum vectors define the required square k-space geometry. After recombination of all the beams at the final beam splitter (BS5), the resulting wavefunction is given by:

\begin{equation}
    \psi_{square} = \psi_{0^\circ}+\psi_{45^\circ}+\psi_{90^\circ}+\psi_{135^\circ}
\end{equation}

\noindent A hexagonal lattice is generated using Configuration 2, in which three of the four interferometric paths contain magnetic prisms with identical magnetic field strengths, but with the field oriented at $0^\circ$, $60^\circ$, and $120^\circ$ relative to the interferometer plane, while the fourth path is blocked. Together, these prisms produce six beams whose transverse momentum vectors form a hexagonal geometry in k-space. The corresponding state at the final beam splitter is given by

\begin{equation}
    \psi_{hexagonal} = \psi_{0^\circ}+\psi_{60^\circ}+\psi_{120^\circ}
\end{equation}

\begin{figure*}
\centering\includegraphics[width=\linewidth]{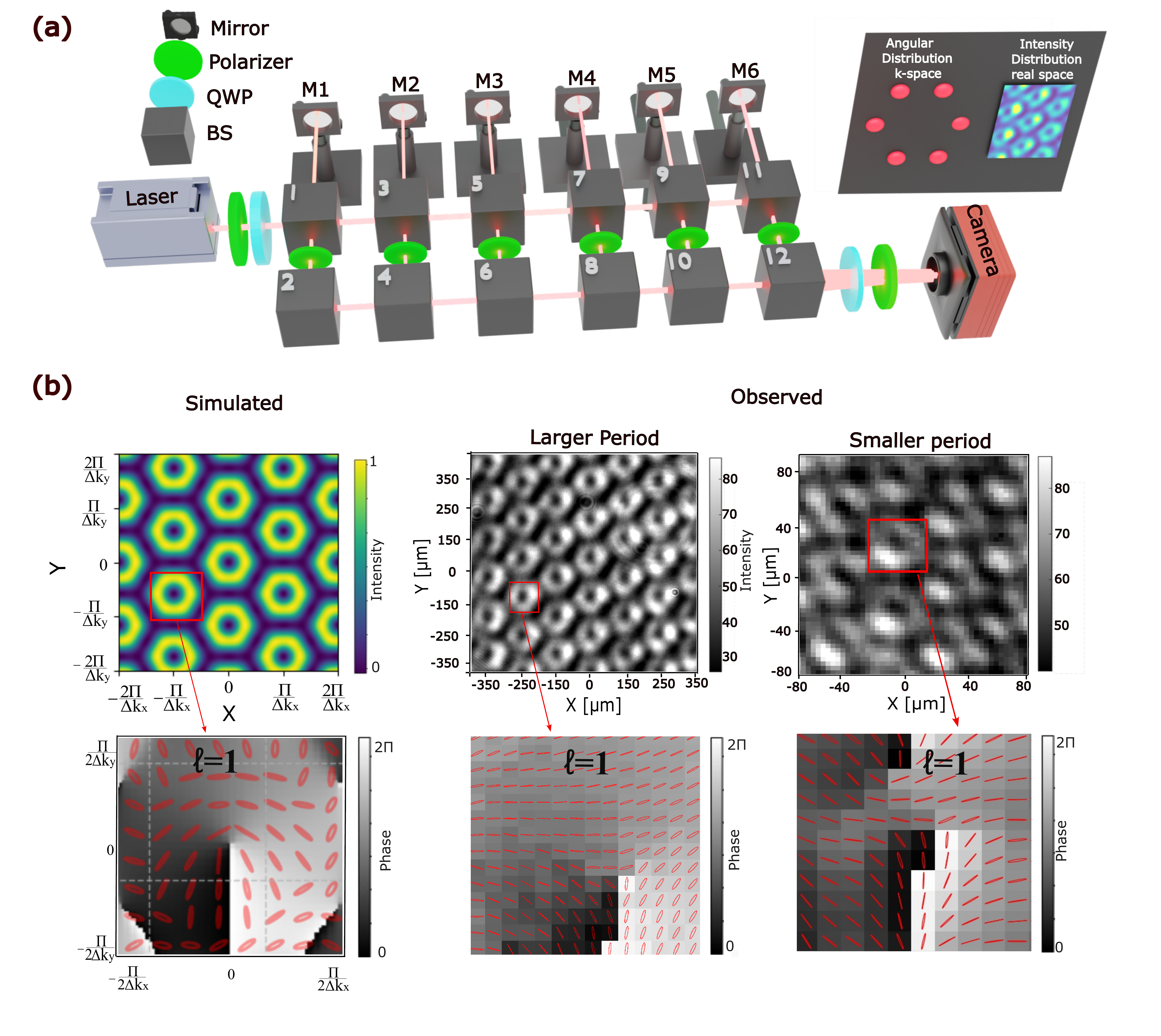}
\caption{(a) Schematic of the experimental setup, implementing the first iteration of the beam-branching operator, for independent control over the phase, polarization, and wave-vector configuration of the interfering beams. A 642 nm circularly polarized laser beam is successively split and recombined through successive stages of an interferometer(BS1–BS12, M1–M6), generating a coherent superposition of six beams. Linear polarizers in each beam path provide independent control of the polarization of each interfering beam, while mirrors mounted on kinematic stages allow adjustment of the wave-vector configuration, relative phase, and spatial overlap. To analyze polarization-dependent interference effects, the output beams are passed through a quarter-wave plate and a linear polarizer before being recorded by a CMOS camera at the imaging plane. (b) Simulated and observed spin–orbit lattices. Column~1: top row shows the simulated intensity profile of six beams following the first iteration of the branching operator, post-selected on left-circular polarization, while the bottom row shows the corresponding overlay of the phase map and polarization ellipse distribution for a unit cell. Columns~2 and~3: top row shows experimentally observed intensity profiles for lattice periods of 158~$\mu$m and 50~$\mu$m, respectively, while the bottom row shows the corresponding phase variations, illustrating an OAM of $\ell = 1$, overlaid with polarization ellipse distributions extracted from the experimental data using Stokes parameter analysis.}

 \label{fig:fig3}
\end{figure*}

 To analyze the resulting two-dimensional intensity and spin distribution, the neutron beam can be directed onto a position-sensitive detector. A spin analyzer before the detector would enable polarization-resolved measurements. By performing a sequence of measurements with different analyzer settings, we can reconstruct the neutron spin-dependent wave function~\cite{sarenac2019generation}. The intensity measured after projection onto a spin-down analyzer is given by
 
\begin{equation}
I = \left| \langle \downarrow_z \vert \psi \rangle \right|^2 .
\end{equation}

 Here, $\psi$ denotes the neutron wavefunction corresponding to either the square ($\psi_{\text{square}}$) or hexagonal ($\psi_{\text{hexagonal}}$) lattice configuration. Fig.~\ref{fig:fig2} shows the intensity profiles for Configurations 1 and 2, yielding square and hexagonal lattices, respectively, with the spin orientation indicated in red.

\section{Optical Implementation}

To generate a hexagonal lattice of $\ell=1$ and $n_r=1$ following the first branching process, we utilized the experimental setup depicted in Fig.~\ref{fig:fig3}a, which allows individual control of the phase, polarization, and angle of each interfering beam, thereby enabling precise tuning of both the structural and unit-cell parameters. 
\subsection{Methods}
A collimated 642 nm laser beam was set into circular polarization by passing it through a polarizer and a quarter‑wave plate. The beam was then successively split and recombined in a cascaded interferometric arrangement consisting of twelve beam splitters (BS1–BS12) and six mirrors (M1–M6), producing six output beams whose interference forms a periodic, structured intensity distribution on the camera screen.

A polarizer was placed in each beam path to set the polarization state of each beam. All mirrors were held using kinematic mounts on the translational stages, allowing precise control of the beams propagation direction, phase delays, and spatial overlap. A CMOS camera was placed in the near field where the beams overlapped, capturing the spatial intensity distribution. Before the camera, a polarizer and a quarter wave plate were installed to allow the selective analysis of polarization-dependent interference effects. The polarization profile was constructed by first capturing polarized images with the analyzer oriented along the \{H, V, D, AD, R, L\} polarization states. From these images, the four Stokes parameters \{S$_0$, S$_1$, S$_2$, S$_3$\} were determined. The four Stokes parameters correspond to the total intensity (S$_0$) and the polarization components along the horizontal/vertical (S$_1$), diagonal/anti-diagonal (S$_2$), and right/left circular (S$_3$) axes of the Poincaré sphere, respectively. The Stokes parameters for each pixel were reconstructed using the following relations: $\{S_0, S_1, S_2, S_3\} = \{I_H + I_V,\ I_H - I_V,\ I_D - I_{AD},\ I_R - I_L\}$. Based on the Stokes parameters, the handedness and eccentricity of the polarization ellipses is computed and plotted in Fig.~\ref{fig:fig3}b (bottom row).

\subsection{Results and discussion}

With the setup shown in Fig.~\ref{fig:fig3}a, we generated a hexagonal lattice, where in each unit cell the spin-orbit state is given by:

\begin{equation}
\Psi(x, y) \approx
\left[\, |R\rangle + e^{i (\phi+\zeta)} \, \sin\!\left(\frac{n_r \pi}{r_c} r\right) |L\rangle \,\right]
\end{equation}

\noindent where $(r, \phi)$ are the cylindrical coordinates, $\zeta$ is a phase offset, and $r_c$ is the radial distance over which the polarization state performs a full rotation on the Poincaré sphere. The periodic beam structure obtained in our experiment establishes the conditions required for the Talbot effect. Therefore, it is important to position the camera at the Talbot distance, given by $z_T = \tfrac{2p^2}{\lambda}$, where the initial periodic intensity pattern is self‑imaged and reappears.

To characterize this state using polarimetry, we can project it onto different polarization bases. For example, projection onto the LCP ($|L\rangle $) yields the intensity profile

\begin{equation}
I(x, y) = \left| \langle L \,|\, \Psi(x, y) \rangle \right|^2
\end{equation}

\noindent with a lattice spacing $p = \frac{2\lambda}{3 \sin\theta}$, as shown in the 2nd and 3rd column of Fig.~\ref{fig:fig3}b (1st row). We find strong agreement between the simulated (Fig.~\ref{fig:fig3}b, 1st column 1st row) and the experimentally observed intensity pattern.  To demonstrate the presence of a lattice of OAM states, the  Stokes phase  $\phi_{12} = \frac{1}{2} \arctan\left(\frac{S_2}{S_1}\right)$ was computed pixel by pixel using the experimentally obtained Stokes parameters. Fig.~\ref{fig:fig3}b (2nd and 3rd columns, 2nd row) presents the resulting phase map for a single lattice site, illustrating an OAM state with $\ell = 1$ through a single $2\pi$ azimuthal phase winding. The ellipticity parameter, $\chi = \tfrac{1}{2}\sin^{-1}\!\left(\tfrac{S_3}{S_0}\right)$, and the azimuth parameter, $\phi_{12} = \tfrac{1}{2}\tan^{-1}\!\left(\tfrac{S_2}{S_1}\right)$, are also calculated at each pixel using the experimentally determined Stokes parameters. Using these parameters, the full state‑of‑polarization (SOP) distribution across the lattice is then determined, as shown in Fig.~\ref{fig:fig3}b (2nd and 3rd columns bottom row). The lattice constant (period) was measured to be 158~$\mu$m (Fig.~\ref{fig:fig3}b, 2nd column), which depends on the wave vector $k$ and angle $\theta$, which describe the tilt of the wave‑vector components ($k_x, k_y$) with respect to the propagation direction. By increasing $\theta$, the lattice period can be consequently reduced, which can be realized by tuning the beam angles through mirrors mounted on kinematic stages, allowing for precise control over the propagation directions of the interfering beams. Using this control, we were also able to generate lattices with a smaller period of approximately 50~$\mu$m, as shown in the third column of Fig.~\ref{fig:fig3}b. Due to the CMOS camera's limited resolution ($5.86~\mu\mathrm{m} \times 5.86~\mu\mathrm{m}$), recorded images appear to be of inferior quality for lattices with short periods.

In addition, in comparison to our earlier work using coherent averaging~\cite{sarenac2018generation}, where the lattice constant obtained was 1.68~mm, the present configuration achieves a significantly smaller period of 50~$\mu$m, representing more than an order-of-magnitude improvement. Similarly, we achieved a period of 4~mm with neutrons~\cite{sarenac2019generation}, which could be reduced by applying the same approach.

\section{Conclusion}

In conclusion, we have proposed and experimentally demonstrated a technique for generating lattices of spin-orbit beams with independent control over lattice symmetry, period, and the OAM and radial numbers associated with each unit cell. Our approach is particularly useful for generating lattices of matter-wave beams, for which we analyzed a feasible experimental configuration. We also report an optical implementation of the technique by generating hexagonal lattices with periods of 158~$\mu$m and 50~$\mu$m and with $\ell = +1$ and $n_r = 1$ in each unit cell. Beyond beam synthesis, the ability to independently tune the internal mode structure and the lattice periodicity may enable structured probes whose spatial modes and lattice scale can be matched to characteristic length scales in a sample, providing new opportunities for mode-selective imaging and materials characterization.

\bibliography{OAM}

\end{document}